\documentclass[journal]{IEEEtran}

\usepackage{orcidlink,float,pgfplots,amsmath,xcolor,soul,graphicx,caption,multirow,makecell,url,siunitx,csquotes}
\usepackage[linesnumbered,ruled,vlined]{algorithm2e}
\hypersetup{hidelinks}

\usepackage{tikz,lipsum}

\usetikzlibrary{positioning,matrix,shapes.arrows,decorations.pathreplacing,calligraphy,fit,arrows.meta}

\definecolor{tudblue100}{HTML}{00305D}
\definecolor{tudblue90}{HTML}{1A456D}
\definecolor{tudblue80}{HTML}{33597D}
\definecolor{tudblue70}{HTML}{4D6E8E}
\definecolor{tudblue60}{HTML}{66839E}
\definecolor{tudblue50}{HTML}{8098AE}
\definecolor{tudblue40}{HTML}{99ACBE}
\definecolor{tudblue30}{HTML}{B3C1CE}
\definecolor{tudblue20}{HTML}{CCD6DF}
\definecolor{tudblue10}{HTML}{E6EAEF}
\definecolor{telekomMagenta}{HTML}{E20074}
\definecolor{tudgreen100}{HTML}{008244}
\definecolor{tudred100}{HTML}{E8412C}
\definecolor{tudorange100}{HTML}{EF7D00}

\usepackage[acronym,shortcuts]{glossaries}
\newacronym{acr}{ACR}{Algebraic Consistency Check Rule}
\newacronym{add}{ADD}{Average Decoding Delay}
\newacronym{arq}{ARQ}{Automatic Repeat Requests}
\newacronym{ber}{BER}{Bit Error Rate}
\newacronym{ct}{CT}{Completion Time}
\newacronym{cs}{CS}{Compressed Sensing}
\newacronym{crc}{CRC}{Cyclic Redundancy Check}
\newacronym{daprac}{DA-PRAC}{Data-aware \acrlong{prac}}
\newacronym{fec}{FEC}{Forward Error Correction}
\newacronym{fprac}{Fly-\acrshort{prac}}{Fly-\acrlong{prac}}
\newacronym{fprnc}{F-PRNC}{Fast Decoding Algorithm for Packet Recovery in \acrlong{nc}}
\newacronym{harq}{HARQ}{Hybrid \acrlong{arq}}
\newacronym{icrc}{ICRC}{Inner \acrlong{crc}}
\newacronym{rlnc}{RLNC}{Random Linear \acrlong{nc}}
\newacronym{nc}{NC}{Network Coding}
\newacronym{ocrc}{OCRC}{Outer \acrlong{crc}}
\newacronym{ppr}{PPR}{Partial Packet Recovery}
\newacronym{pprcs}{PPR-CS}{\acrlong{ppr} with Compressed Sensing}
\newacronym{prac}{PRAC}{Packetized Rateless Algebraic Consistency}
\newacronym{snr}{SNR}{Signal-to-Noise Ratio}
\newacronym{snc}{SNC}{Sparse \acrlong{nc}}
\newacronym{spac}{SpaC}{Simple packet Combining}
\newacronym{sprac}{S-PRAC}{Segmentized \acrlong{prac}}

\pgfplotsset{compat=newest}
\usepgfplotslibrary{patchplots}

\makeatletter
\let\ftype@table\ftype@figure 
\makeatother

\newcommand{\floor}[1]{\left\lfloor #1 \right\rfloor}

\begin{document}
\title{\acrshort{fprac}: Packet Recovery for \acrlong{rlnc}}

\author{Hosein~K.~Nazari\,\orcidlink{0000-0003-1632-4782},
        Stefan~Senk\,\orcidlink{0000-0003-0745-2264},
        Peyman~Pahlevani\,\orcidlink{0000-0001-5918-7250},
        Juan~A.~Cabrera\,\orcidlink{0000-0002-7525-2670},
        Frank~H.~P.~Fitzek\,\orcidlink{0000-0001-8469-9573}
\thanks{H.~K.~Nazari, S.Senk, Juan~A.~Cabrera, and F.~H.~P.~Fitzek are with Technische Universität Dresden (TUD), Deutsche Telekom Chair of Communication Networks}%
\thanks{F.~H.~P.~Fitzek are with the Cluster of Excellence “Centre for Tactile Internet with Human-in-the-Loop” (CeTI) of TUD}
\thanks{P.~Pahlevani is with Department of Electronic Systems, Aalborg University, Aalborg, Denmark}%

}

\markboth{}
{\acrshort{fprac}: Packet Recovery for \acrlong{rlnc}}

\maketitle

\begin{abstract}
\acrfull{nc} is a compelling solution for increasing network efficiency.
However, it discards corrupted packets and cannot achieve optimal performance in noisy communications.
Since most of the information in corrupted packets is error-free, discarding them is not the best strategy.
Several packet recovery techniques such as \acrshort{prac} and \acrshort{sprac} were proposed to exploit corrupted packets.
Yet, they are slow and only practical when the packet size is small and communication channels are not very noisy.
We propose a packet recovery scheme called \acrshort{fprac} to address these issues.
\acrshort{fprac} exploits algebraic relations between a group of coded packets to estimate their corrupted parts and recovers them.
Unlike previous schemes, \acrshort{fprac} can recover coded packets at the intermediate node without decoding them.
We have compared \acrshort{fprac} against \acrshort{sprac}.
Results show when the bit error rate ($\epsilon$) is $10^{-4}$
, \acrshort{fprac} outperforms \acrshort{sprac} by two folds for a payload of \SI{900}{\byte}.
In two-hop communication with $\epsilon$ = $10^{-4}$ and a payload size of \SI{500}{\byte}, by enabling the recovery in the intermediate node, \acrshort{fprac} reduces transmissions by 16\%.
In a \acrfull{snc} scenario, with two non-zero elements in the coefficient vectors and a payload of \SI{800}{\byte}, there is a reduction by 31\% on average for decoding delay.
\end{abstract}

	
%

\section{Introduction}
\label{sec:introduction}

\IEEEPARstart{P}{acket} errors in wireless communications are common and inevitable.
These errors degrade the throughput and cause delays in communication systems.
Generally, to provide a reliable communication, a receiver needs to check the integrity of received information.
For this purpose, an error detection code such as a \gls{crc} is associated with every packet.
Subsequently in this article, we refer to packets with erroneous bits as \textit{partial packets}.
\gls{crc} helps to detect partial packets and refutes their integrity.
Conventionally, to obtain reliable communication in the data-link layer, techniques such as \gls{arq} and \gls{fec} are used.
In \gls{arq}, the receiver discards partial packets and issues a retransmission request.
With \gls{fec} codes, an encoder transforms $k$~bits of the original packet into $n$~coded bits at the transmitter side, where $n > k$.
Then, a receiver can use the additional redundant bits to recover partial packets.
Moreover, \gls{harq}~\cite{b1} uses a combination of both techniques to further improve reliability.
Besides their advantages and wide adoption in commercial devices, the aforementioned solutions are sub-optimal.
For instance, \gls{arq} is unsuitable for communications with a low \gls{snr}.
In addition, \gls{fec} codes have limited correction capabilities and merely perform well within a specific range of \gls{snr}~\cite{FECSNR1,FECSNR2}.
	

In~\cite{NC}, Ahlswede et al. propose a novel concept called \gls{nc} that has arisen as a possible solution to resource utilization and gaining higher reliability~\cite{NCReliable1,NCReliable2}.
\gls{rlnc}, introduced in~\cite{RLNC}, is a decentralized version of \gls{nc}.
In \gls{rlnc}, data is split into multiple generations.
A generation contains $g$ original packets, and each original packet comprises several equally-sized symbols.
\gls{rlnc} randomly combines original packets to form coded packets.
Once a receiver collects a sufficient number of coded packets, it decodes them.
However, \gls{rlnc} suffers from high computational complexity and long decoding delay.
Therefore, \gls{snc}~\cite{SNC} has been proposed to address these issues.
However, \gls{rlnc} and \gls{snc} can only decode \textit{error-free} packets while discarding partial received packet fragments.

The authors in~\cite{ziptx} show, that roughly $55\%$ of all received packets in the IEEE~802.11a/b/g environment are partial.
In~\cite{Waste}, the authors report that between $5\%$ to over $50\%$ of received packets IEEE~802.11g/n networks are partial packets, while only between $2$ to $10$ percent of received bits are erroneous.
Yet, researchers demonstrate in~\cite{PartialWaste} that up to $95\%$ of a partial packet consists of error-free information.
Thus, methods that discard partial packets are sub-optimal solutions because they remove useful and error-free information.
	
	
The first \gls{ppr} method is proposed in~\cite{firstPPR}.
A variety of methods, such as~\cite{crosslayer1}, use cross-layer information , i.e., from the physical layer, to estimate the location of errors.
In~\cite{SOFT}, a novel architecture called \textit{SOFT} receives a measurement of confidence from the physical layer for each received bit to estimate the error locations.
A link-layer retransmission technique called \textit{PP-ARQ}~\cite{PPARQ} first finds corrupted parts of received packets and then only asks for the retransmission of these particular parts.
Since passing such information violates the current layered network protocols and standards, it requires modification at the hardware level.
Similarly, in segment-based \gls{ppr}~\cite{segmentBasedOne,segmentBasedTwo}, each packet is segmented, so that a recipient must only request retransmissions for corrupted segments.
In wireless communication, a technique called \textit{spatial diversity} suggests dedicating more than one antenna to either receiving or transmitting data.
Several schemes such as \gls{spac}~\cite{Spac} take advantage of spatial diversity in their \gls{ppr} scheme.
For instance, \gls{spac} recovers partial packets by comparing and combining them, when at least two versions of a packet have been received.
	
Currently, partial packets are disregarded in different implementations of \gls{nc}, such as \gls{rlnc} and \gls{snc}.
But, \gls{ppr} solutions can help to achieve higher reliability and throughput as a complementary strategy.
In~\cite{PPR_CS}, the authors suggest a \gls{ppr} scheme for \gls{rlnc} called \acrshort{ppr}-\acrshort{cs} that relies on data techniques such as \gls{cs}.
\acrshort{ppr}-\acrshort{cs} does not require any cross-layer information and attempts to recover partial packets by solving a set of sparse recovery problems.
Furthermore, Angelopoulos et al. introduce a novel \gls{ppr} scheme called \gls{prac} in~\cite{PRAC} that only relies on the inter-packet algebraic consistency in \gls{rlnc}, and does not require any information from the physical layer.
\gls{prac} proposes a method called \gls{acr} to estimate corrupted locations of partial packets, and then uses an iterative correction algorithm to recover the packets.
Besides its benefits, \gls{prac} suffers from high computational complexity.
Also, it is only applicable, where the size of packets is small, or the level of noise in a channel does not exceed a certain threshold~\cite{Waste}.
To address the high resouce demand of \gls{prac}, several \gls{ppr} methods such as~\cite{DAPRAC,SPRAC,F-PRNC} have been proposed taking advantage of \gls{acr} to provide an estimation of corrupted parts.
However, \gls{ppr} schemes based on \gls{acr} suffer from common drawbacks:
One of the most beneficial advantage of \gls{rlnc} compared to other coding techniques is the ability to recode packets at intermediate nodes~\cite{ho2008}, with accompanying subsequent gains.
But, these methods only recover partial packets at the receiver, intermediate nodes can not use partial packets for recoding, yet.
Further, the performance of such methods becomes worse as the number of original packets or symbols increases.
And in some cases, the \gls{acr} even fails to detect all corrupted parts of partial packets, which we are referring to as a \textit{false-positive event} in the rest of the article.
As a consequence, some partial packets may not be recovered without detection.
Our measurements also show, the probability of false-positive events can increase for larger generations.

\subsection{Our Contribution}
\label{sec:intro-contribution}
This article proposes a \gls{ppr} scheme called \gls{fprac} for \gls{nc}-based communication.
\gls{fprac} helps to reduce the number of retransmissions and thus communication overhead.
\gls{fprac} can improve throughput where the cost of retransmission due to channel condition, low transmission rate, and packet sizes is high.
This method aims to enhance recovery speed and throughput for different ranges of packet and generation sizes.
	
\gls{fprac} at the transmitter side divides 
coded packets into $s$~equal-sized segments, and then computes a checksum for each coded packet.
After transmitting $R-1$~coded packets, the sender transmits a linearly dependent packet by combining the last $R-1$~transmitted coded packets.
This dependent packet enables the receiver to estimate the corrupted parts of this group of packets and to begin the recovery process.
	
The main advantages of \gls{fprac} are summarized as follows:
\begin{itemize}
	\item This method can recover partial packets at intermediate and receiver nodes.
	Thus, an intermediate node can recover partial packets and use them to form recoded packets in order to gain more efficiency.
	\item Compared to methods based on \gls{acr}, \gls{fprac}'s probability of false-positive events in the recovery process is lower, and as a consequence, \gls{fprac} can recover more partial packets.
	\item A version of the proposed method is designed for \gls{snc} communication which exploits dependent packets to recover partial packets.
	\item \gls{fprac} provides a more precise estimation of error locations, enabling it to recover partial packets with fewer operations than methods based on \gls{acr}.
	\item \gls{fprac} provides an estimation and does a correction process for each group of $R$~packets separately.
	Consequently, \gls{fprac} starts recovery earlier and thereby reduces the completion time of the recovery process.
	\item The proposed method can be easily adapted to different error rates and channel conditions by changing parameters such as~$R$.
\end{itemize}

We have conducted extensive simulations \gls{fprac}'s performance and compared it with one of the state-the-art solutions \gls{sprac}.
The results illustrate, \gls{fprac} dramatically enhances the communication efficiency in poor channel conditions.
For instance, in an \gls{rlnc} communication, for a \gls{ber} of $5 \times 10^{-5}$, the completion time can be decreased by up to $4.7$~times.
We have compared \gls{fprac} with \gls{rlnc} and \gls{sprac} in terms of goodput.
Goodput refers to the amount of data transferred and decoded by the receiving node per second.
For large payload sizes such as \SI{900}{\byte}, the proposed method can increase the goodput by up to $3.8$~times.
Furthermore, the simulation shows that recovery at intermediate nodes reduces overall transmissions by up to $16\%$.
In very noisy communication ($\gls{ber} = 5 \times 10^{-5}$) and for a generation of $100$~packets, it can recover $14$~times more partial packets compared to \gls{sprac}.
Additionally, in an \gls{snc} communication and for a \gls{ber} of $10^{-4}$, utilizing \gls{fprac} leads to a reduction of up to $47\%$ on average in decoding delay.
	
\subsection{Paper Organization}
\label{sec:intro-paper-organization}
This article is organized as follows:
In \autoref{sec:background-related-work}, state-of-the-art error-controlling and \gls{ppr} schemes are reviewed.
We introduce \gls{fprac} and its encoding and recovery procedures in \autoref{sec:system-model}.
In \autoref{sec:recoding-at-intermediate-nodes}, we propose a method for recovering partial packets, and in \autoref{sec:packet-recovery-snc}, we anaylze the benefits of the proposed method in \gls{snc}-based communications.
Afterwards, we discuss the impact of different parameters on the recovery performance in \autoref{sec:parameter-analysis}.
In \autoref{sec:simulation-results}, we present our simulation results, and finally, the concluding remarks are given in \autoref{sec:conclusion}.
	
In addition, \autoref{table:notation} provides descriptions for some of the notations used in this manuscript.

\begin{table*}[ht]
\begin{center}
    \renewcommand{\arraystretch}{1.2}
    \caption{Symbols and notations used in this article.}
    \label{table:notation}
    \begin{tabular}{|l|l|}
        \hline
        \textbf{Notation}        & \textbf{Description}                                                        \\ \hline
        GF($2^{q}$)     & Finite Field with size of  $2^{q}$                                                   \\ \hline
        $\epsilon$      & \acrlong{ber}                                                                        \\ \hline
        $A_{w}^{i}$     & Number of valid code words with Hamming weight of $w$                                \\ \hline
        $b$             & Number of bits in a symbol                                                           \\ \hline
        $s$             & Number of segments in a coded packet                                                 \\ \hline
        $l$             & Number of symbols in a packet                                                        \\ \hline
        $R$             & Number of packets in a dependent group                                               \\ \hline
        $ p_{i}$        & $i$th original packet                                                                \\ \hline
        $ p'_{i} $      & $i$th coded packet                                                                   \\ \hline
        $ s_{i,j} $     & $j$th original symbol of $i$th original packet                                       \\ \hline
        $ s'_{i,j} $    & $j$th coded symbol of $i$th coded packet                                             \\ \hline
        $ C $     		& Coefficient matrix                                       						       \\ \hline
        $ P''$          & Transmission matrix                             									   \\ \hline
        $ M $     		& Reception matrix in \acrlong{rlnc}                          						   \\ \hline
        $ M' $     		& Reception matrix in \acrlong{snc}                     							   \\ \hline
        $g$             & Generation size                                                                      \\ \hline
        $N_{p}$         & Number of bits in a segment including associated \acrlong{icrc}.                     \\ \hline
        $B_{valid}$     & The buffer of valid packets.                                                         \\ \hline
        $B_{invalid}$   & The buffer of invalid packets.                                                       \\ \hline
        $P_{h}^{s}$     & The probability of receiving an error-free symbol                                    \\ \hline
        $P_{h}^{c}(R)$  & The probability of receiving an error-free column of $R$ coded symbols               \\ \hline
        $E_{ic}$        & The expected number of inconsistent columns in the estimation process                \\ \hline
        $P^{(1)}_{fpe}$ & The probability of false-positive in a round of estimation process                   \\ \hline
        $P_{un}$        & The probability of undetected error in a segment                                     \\ \hline
        $P_{sr}$        & The probability of successful recovery of a segment in correction process            \\ \hline
        $P_{e}^{i}$     & The probability of $i$ symbols being erroneous in a round of estimation              \\ \hline
        $P_{fe}^{(1)}$  & The probability of failure caused by estimation in one round of estimation process   \\ \hline
        $P_{fe}^{(l)}$  & The probability of failure caused by the estimation process for a packet             \\ \hline
        $P_{fpc}^{(1)}$ & The probability of a false-positive event during correction trials for a segment     \\ \hline
        $P_{fc}^{(s)}$  & The probability of failure caused by false-positive events in the correction process \\ \hline
        $P_{f}$         & The probability of failure in recovering a packet in \acrshort{fprac}             \\ \hline
    \end{tabular}
\end{center}
\end{table*}

\section{Background and Related Work}
\label{sec:background-related-work}
Maintaining reliable communication over noisy networks and utilizing error-free information of partial packets to improve overall communication performance has attracted extensive attention.
This section will go through some of the most important methods.

\subsection{Automatic Repeat Request and Forward Error Correction}
\label{sec:arq-fec}
Traditionally, \gls{fec} and \gls{arq} are widely being used to combat the erroneous channel's challenges and provide higher reliability.
\gls{fec} appends a constant overhead to packets and provides limited correction capabilities.
Due to fast-varying and unstable channel quality in wireless networks, it is necessary to consider the worst-case scenario to achieve high reliability.
Considering the worst-case scenario leads to inefficient utilization of network resources.
Moreover, acknowledgment packets can provide reliability in some conditions, although it seems impractical in poor channel conditions and broadcast scenarios.

\gls{harq} approaches, e.g., \cite{HARQ1}, combine \gls{fec} and \gls{arq} to propose a more flexible method.
\gls{harq} aims to recover a portion of partial packets using error correction codes to reduce the overhead caused by retransmissions.
\gls{harq} approaches suffer from the same challenges as \gls{arq} methods in broadcast networks; when different receivers lose different packets, a retransmitted packet is only beneficial to receivers missing that packet.
In contrast, in \gls{rlnc} based communication, a coded packet can be useful to more receivers \cite{HARQvsRLC}.


\subsection{Partial Packet Recovery using Physical Layer Information}
\label{sec:ppr-using-phy-info}
Several partial packet recovery techniques use physical layer information to estimate erroneous portions of packets.
For instance, approaches based on soft-decoding use a confidence measure provided by the physical layer to estimate the location of erroneous bits.
These measurements are sent to higher layers, where the receiver issues a request to retransmit those bits.
Besides its advantages, since passing such information violates the current layered network protocols and standards, it also requires modification at the hardware level~\cite{PPRreview}. 
	
\subsection{Partial Packet Recovery in Network Coding}
\label{sec:ppr}
A variety of packet recovery techniques are proposed for network coding communications.
These methods use the relation between coded packets to identify error locations and correct them.
Since the recovery process of these methods is time-consuming, they are beneficial when the retransmission cost in terms of delay and energy is high. 
	
\gls{pprcs}~\cite{PPR_CS} relies on data processing techniques and works transparently from the sender.
\gls{pprcs} uses a systematic RLC encoder and decoder, exploits the coding matrix's algebraic properties, and employs compress sensing to recover partial packets.
However, it relies on the physical layer to correct most bit errors using \gls{fec} techniques and assumes that errors are sparse, random, and independently distributed among different packets. 
For \gls{pprcs}, smaller generation sizes are suggested in poor channel conditions to decrease the sparsity order of errors. 
 

\gls{prac}~\cite{PRAC} is the first approach that suggests exploiting the algebraic consistency of coded packets to estimate the location of erroneous symbols.
It begins the recovery process after collecting $g$ or more packets.
First, it selects $g+1$ packets from both partial and error-free packets.
Then, it uses $g$ packets in the decoding process.
Next, it uses this information to recode the other packet; and finally, by comparing the estimated and received versions, it estimates error locations.
This process runs over each column separately and is called \acrfull{acr}.
In case of not verifying a column by an \gls{acr}, the bits involved in the column are permuted, and the \gls{acr} is repeated.
This exhaustive search and permutation continue until the \gls{acr} verifies that the estimated and received versions of symbols are the same.
	
This technique does not need any overhead over packets and boosts the throughput in good channel conditions.
Still, \gls{prac} suffers from a slow recovery process, making it impractical for large generation and packet sizes in poor conditions with higher error rates.
Additionally, the recovery begins only after receiving the $(g+1)$th packet which causes a considerable delay.
	
Several schemes are proposed to address the high computational complexity of \gls{prac}.
\gls{daprac}~\cite{DAPRAC} suggests decoding packets using different sets of either partial or valid coded ones and denotes them as predecoded packets.
Then, it selects one set of predecoded packets and tries to detect and recover partial packets using other sets.
One of the main issues of \gls{daprac} is the time-consuming process of providing different groups of pre-decoded packets, mainly when the generation or payload size is large.
Further, authors of~\cite{SPRAC} proposed \gls{sprac} with focusing on highly noisy channel conditions.
They suggest splitting packets into different segments and dedicating a \acrshort{crc}-8 to each one.
\gls{sprac} carries the recovery process for each invalid segment separately and uses \gls{acr} to estimate the location of errors within it.
Although \gls{sprac} can boost recovery speed, it suffers from falsely verified segments during recovery.
Further, \gls{fprnc}~\cite{F-PRNC} is also a \gls{ppr} scheme for \gls{rlnc} that suggests splitting coded packets into $k$ coded symbols and later adding $R$ redundancy symbols of the same size by linearly combining $k$ coded ones.
On the decoder side, to recover partial packets, it seeks $k$ error-free symbols among either redundancy or coded ones, and uses \gls{acr} to boost its recovery process.
However, it is not suitable for large payload sizes since increasing the number of symbols slows its recovery process, and also splitting a packet into a fewer number of larger symbols wastes bandwidth.

\section{System Model}
\label{sec:system-model}
Here, we discuss the encoding and recovery processes of the proposed method.
	
        %
        %
        %
	
\subsection{Encoding Process}
\label{sec:sm-encoding-process}
In the encoding process, the transmitter encodes $g$ original packets $\{ {p_{1}},{p_{2}},\ldots,{p_{g}}\}$.
We refer to $g$ as the generation size.
Each packet $p_{i}$ comprises $l$ symbols $\{{s_{i,1}},{s_{i,2}},\ldots,{s_{i,l}}\}$ where each symbol is an element of $GF(2^q)$.

The encoder generates coded packets in groups.
Each group contains $R$ 
coded packets and is called a dependent group.
In a dependent group, one coded packet is linearly dependent.
To generate a dependent group, the encoder forms a matrix containing $g$ original packets denoted by $P_{(g \times l)}$.
Then, it generates a coefficient matrix denoted by $C_{(R \times g)}$.
Elements of the first $R-1$ rows of $C_{(R \times g)}$ are selected randomly from a finite field $GF(2^q)$.
The elements of the last row of $C_{(R \times g)}$ are computed by linearly combining the first $R-1$ rows of $C_{(R \times g)}$ using addition in $GF(2^q)$.
Then $C_{(R \times g)}$ is multiplied by  $P_{(g \times l)}$ to form a transmission matrix which is denoted by $P'_{(R \times l)} $. 
\begin{equation}
	P'_{(R \times l)} = C_{ (R \times g)} \times P_{(g \times l)}.
\end{equation}
Each row of  $P'_{(R \times l)} $ contains coded symbols of a coded packet.
Since decoding a generation of $g$ packets requires at least $g$ linearly independent coded packets, the encoder generates at least $\frac{g}{R-1}$ dependent groups.

\begin{figure}[t!]
    \begin{center}
        \includegraphics[width=\columnwidth]{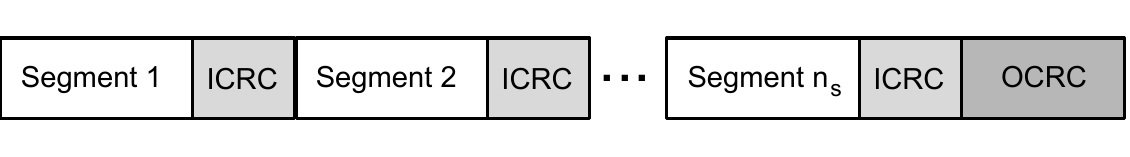}
        \caption{Format of a coded packet.}
        \label{fig:coded-packet-structure}
    \end{center}
\end{figure}

Each coded packet $p'_{i}$ contains $l$ coded symbols $\{ s'_{i,1},s'_{i,2},...,s'_{i,l}\}$ and is divided into $s$ equally-sized segments where it is assumed that $l/s$ is an integer value.  
Then, for each segment, an \gls{icrc} is computed over the segment, and appended to the coded packets.
The \gls{ocrc} is computed over all coded symbols, and is appended to the packet.
Therefore, each $p'_{i}$ contains $s$ segments, $s$ \glspl{icrc} and one \gls{ocrc}.
\autoref{fig:coded-packet-structure} shows the format of a coded packet.
Each coded packet conveys the associated set of coefficients, or the transmitter synchronizes its random coefficient generators with the receiver.
	
\subsection{Recovery Process}
\label{sec:sm-recovery-process}
Upon receiving a packet, its consistency is checked using the associated \gls{ocrc}.
If the \gls{ocrc} verifies a packet, it is categorized as a valid packet and inserted to $B_{\textit{valid}}$ buffer.
Afterwards, it is passed to the decoder to obtain partially decodable symbols.
On the other hand, if \gls{ocrc} refutes a packet, the packet is categorized as an invalid one and inserted into $B_{\textit{invalid}}$.
	
In a dependent group, one packet is linearly dependent; thus, after receiving a dependent group, if $B_{\textit{invalid}}$ contains two or more invalid packets, the recovery begins.
The recovery comprises two subsequent phases: 1)~error location estimation, and 2)~correction by permutation.

In this research, it is assumed that only the coded symbols can have errors in received packets.
Several techniques can be used to achieve error-free coefficients.
In a network without recoding, the sender and receiver can synchronize their random coefficient generators, and there is no need to transfer coefficients at all.
Furthermore, in the case of communication with recoding, methods such as~\cite{reduceCoeffSize} could be employed to significantly reduce the size, and thus the probability of error occurrence in coefficients.

%
	
\subsubsection{Estimation of Error Locations}
\label{sec:estimation-error-locations}
\begin{figure*}
    \begin{center}
        \begin{tikzpicture}[scale=1,
                            transform shape,
                            decoration={calligraphic brace,
                                        mirror,
                                        amplitude=6pt,
                                        raise=2pt
                                        }
                            ]
            \node[single arrow,
                  anchor=west,
                  draw=black,
                  right color=tudblue100,
                  left color=tudblue10,
                  minimum height=18cm,
                  draw opacity=0.5,
                  fill opacity=0.8,
                  text opacity=1.0
                  ] (bottomArrow) at (0,-0.5) {\small Error Estimation in Recipient};
            \matrix[matrix of math nodes,
                    row sep=0.15cm,
                    left delimiter=(,
                    right delimiter=)
                    ] (coeff) at (1.5,3)
            {
                c & c & c\\
                c & c & c\\
                c & c & c\\
            };
            \matrix[matrix of math nodes,
                    row sep=0cm,
                    left delimiter=(,
                    right delimiter=)
                    ] (recvSym) at (3.5,3)
            {
                \node[tudgreen100] {s'}; & \node[tudgreen100] {s'};\\
                \node[tudgreen100] {s'}; & \node[tudred100] {s'};\\
                \node[tudgreen100] {s'}; & \node[tudgreen100] {s'};\\
            };
            \node[text width=2cm,
                  align=center
                  ] (coeffText) at (1.5,4.25) {\footnotesize Coefficient Matrix};
            \node[text width=2.5cm,
                  align=center
                  ] (recvSymText) at (3.5,4.25) {\footnotesize Received Coded Symbols};
            \node[anchor=west,
                  align=left,
                  text width=4cm
                  ] (1) at (0.5,1.25) {\footnotesize (1) Once a packet is received, its data is saved in coefficient and coded symbol matrices.};
            \node[single arrow,
                  anchor=west,
                  draw=black,
                  right color=tudblue100,
                  left color=tudblue10,
                  minimum height=4cm,
                  draw opacity=0.5,
                  fill opacity=0.8
                  ] (bottomArrow) at (5.25,3) {};
            \node[anchor=west,
                  align=left,
                  text width=4cm
                  ] (2) at (5.25,3.75) {\footnotesize (2) After receiving $R$ packets, the matrices are concatenated.};
            \node[anchor=west,
                  align=left,
                  text width=4cm
                  ] (3) at (5.25,2) {\footnotesize (3) The reduced echelon form of the resulting matrix from step 2 is calculated.};
            \matrix[matrix of math nodes,
                    anchor=west,
                    row sep=0.15cm,
                    left delimiter=(,
                    right delimiter=)
                    ] (matrix2) at (10.75,3)
            {
                c & c & c & \, & s' & s' \\
                c & c & c & & s' & \node[tudred100] (matrix2-2-6) {s'}; \\
                0 & 0 & 0 & & 0 & \emptyset \\
            };
            \draw (12.25,3.95) -- (12.25,2.05);
            \draw[decorate,
                  thick
                  ] (12.2,4) -- (10.75,4) node[midway,
                                               yshift=2em,
                                               xshift=-0.5em,
                                               text width=2cm,
                                               align=center
                                               ]{\footnotesize Coefficient Section};
            \draw[decorate,
                  thick
                  ] (13.5,4) -- (12.3,4) node[midway,
                                              yshift=2em,
                                              xshift=0.5em,
                                              text width=2cm,
                                              align=center
                                              ]{\footnotesize Coded Symbols Section};
            \node[fit=(matrix2-1-6)(matrix2-3-6),
                  draw=tudred100,
                  inner sep=0pt,
                  rounded corners=6pt,
                  dash dot,thick
                  ] (box1) {};
            \node[fit=(matrix2-3-1)(matrix2-3-6),
                  fill=tudblue100,
                  fill opacity=0.2,
                  inner sep=0pt,
                  rounded corners=6pt
                  ] (box2) {};
            \node[fit=(matrix2-2-6),
                  fill=tudred100,
                  fill opacity=0.3,
                  inner sep=2pt,
                  rounded corners=6pt
                  ] (box3) {};
            \draw[thick,
                  arrows={Latex-[width=10pt,
                                 length=10pt
                                 ]
                          }
                  ] (box3.north east) to [bend right=-30] (14.75,3.5) 
                  node[anchor=north west,
                       xshift=-2em
                      ] {\footnotesize Erroneous Symbol};
            \draw[thick,
                  arrows={-Latex[width=6pt,
                                 length=6pt
                                ]
                          }
                  ] (box2.west) to [bend right=45] (10,1.5) 
                  node[anchor=north west,
                       text width=3.75cm,
                       align=left,
                       yshift=-0.25em,
                       xshift=-5em
                       ] {\footnotesize (4) After step 3, all values in one of the rows in the coefficient section will be zero.};
            \draw[thick,
                  draw=tudred100,
                  arrows={-Latex[width=6pt,
                                 length=6pt,
                                 tudred100
                                 ]
                          }
                  ] (box1.south east) to [bend left=45] (14.5,1.75) 
                  node[anchor=north west,
                       text width=5.5cm,
                       align=left,
                       xshift=-5.5em,
                       yshift=-0.25em
                       ] {\footnotesize (5) In the coded symbols section of the row discussed in step 4, the non-zero element indicates that there is at least one erroneous coded symbol within the column.};
        \end{tikzpicture}
        \caption{
        An overview of the error location estimation process.
        In this example, three packets are sent, one of which has a coefficient vector that is linearly dependent on the other two.
        Here, $c$ and $s'$ are referring to coefficients, and coded symbols, respectively.
        Furthermore, the red color refers to an erroneous symbol, and $e$ refers to a non-zero value which shows the location of the erroneous column.
        For simplification, each coded packet contains one segment with two coded symbols.
        One of these elements becomes corrupted during transmission. 
        }
        \label{fig:error-location-estimation-process}
    \end{center}
\end{figure*}
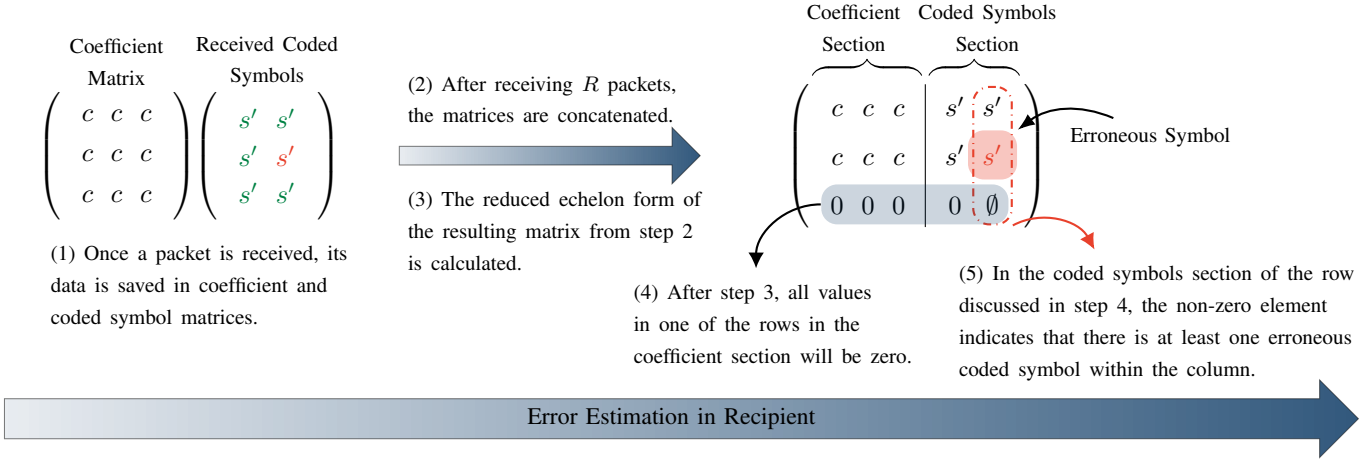

\gls{fprac} uses dependent packets to estimate error locations in each dependent group separately.
This enables \gls{fprac} to provide a precise estimation of error locations.
	
In this phase, a reception matrix is generated denoted by $M$.
Each row of $M$ contains coefficients and symbols of a coded packet; the first $g$ elements are coefficients followed by $l$ coded symbols.
Since one of the packets is linearly dependent, after performing Gaussian elimination over $M$, in one of the rows, all first $g$ elements must be zero.
If coefficients of a coded packet are zero, the resulting coded symbols must be zero.
Therefore, any non-zero element in this row indicates that at least one of the coded symbols in that column is erroneous.
Next, the process generates a vector called $V_{\textit{Broken}}$ including the index of detected inconsistent columns.

\subsubsection{Correction by Permutation}
\label{sec:correction-by-permutation}
Upon estimation process, \gls{fprac} takes advantage of \glspl{icrc} and carries out correction process for each segment separately.
This separation enables \gls{fprac} to reduce the maximum number of correction trials, especially for packets with large payload sizes.
The operation begins for an invalid segment by permuting the value of symbols whose indices are involved in $V_{\textit{Broken}}$.
After each permutation, a \gls{crc} is executed.
These permutations continue until the \gls{icrc} verifies the segment. 
	
To find the answer with fewer permutations, we start with the most probable ones.
Therefore, we begin permutations with minimum Hamming distance from the received version of the segment.
In some cases, you can consider a limit on the number of permutations to reduce the complexity of this process.
	
When all of the invalid segments of a partial packet are verified by their associated \glspl{icrc}, the integrity of the entire packet is checked using \gls{ocrc}.
If the \gls{ocrc} verifies the packet, the recovered version of it is inserted to $B_{\textit{Valid}}$; otherwise, the receiver discards the packet.
Furthermore, since one of the packets in each group of $R$ packets is dependent, the recovery stops whenever $R-1$ packets in the group are recovered or are among valid ones.

Once the receiver obtains $g$ linearly independent valid coded packets, the recovery process terminates, and the receiver can decode all coded packets using the \gls{rlnc} decoder.

\section{Recoding at Intermediate Nodes}
\label{sec:recoding-at-intermediate-nodes}
The main advantage of network coding compared to other coding techniques is the ability to recode packets at intermediate nodes \cite{ho2008}.
If a partial packet combines with other packets to form a recoded one, the obtained recoded packet would be erroneous.
In addition, an error-free version of coded symbols is necessary to compute \gls{crc} checksums for a packet.
If an intermediate node cannot recover partial packets, it would have only two options: discarding partial packets or forwarding them without recoding. 

Moreover, if an intermediate node forwards partial packets, the total number of partial packets passing through several noisy links would be significant.
Forwarded partial packets will contain more corrupted bits, and consequently, more correction trials are required to recover them.  

Therefore, recovering partial packets at the intermediate nodes can help to reduce the recovery process's duration, and enable us to use recovered partial packets in the recoding procedure.

We consider two buffers for valid and invalid packets at the intermediate node.
Once a valid packet arrives, the intermediate node inserts it into valid packets' buffer ($B_{\textit{valid}}$).
Later, the intermediate node generates and transmits a recoded packet using packets in $B_{\textit{valid}}$.
In case of receiving an invalid packet, the intermediate node inserts the packet into invalid packets' buffer ($B_{\textit{invalid}}$).
Once all packets belonging to a dependent group are received, the estimation and correction processes begin.
Next, for each recovered packet, the intermediate node generates a recoded packet using the recovered packet and packets in $B_{\textit{valid}}$, and transmits it.
After this process, the intermediate node discard packets from both $B_{\textit{invalid}}$ and $B_{\textit{valid}}$. 
	
	
%
    %
    %
    %
	
\section{Packet Recovery for Sparse Network Coding}
\label{sec:packet-recovery-snc}
Here, we propose a version of \gls{fprac} for \acrfull{snc}.
One of the main advantageous features of \gls{snc} is partial decoding which means decoding some of the packets before receiving the entire packets of a generation.
This feature is beneficial to a range of applications, such as real-time video transmission.
The main difference between \gls{snc} and \gls{rlnc} is the sparsity level of coefficient vectors.
While \gls{rlnc} chooses the elements of the coefficient vectors at random from a finite field, in \gls{snc}, most of the elements in a coefficient vector are zero.
Apart from its benefits, this feature leads to the generation of a significant number of dependent packets~\cite{zareisncfeature}.

Here, we propose a version of the Fly-\gls{prac} for \gls{snc}-based communications that take advantage of all dependent packets in its recovery process.
In the following, the encoding and recovery processes of this version of Fly-\gls{prac} are discussed.
	
	
\subsection{Encoding Process}
\label{sec:snc-encoding-process}
The encoding processes of \gls{fprac} for \gls{snc} and \gls{rlnc} are similar.
The only difference between these two in the encoding procedure is the generation of coefficient vectors.
Regarding the \gls{snc} scheme, coefficient vectors are $w$-sparse.
It means only $w$ elements of a coefficient vector are non-zero, and their positions are selected randomly.
Also, each of the $w$ non-zero elements is randomly and independently selected from $GF(2^{q})- \{0\}$. 
The rest of the encoding process is similar to the encoding process of \gls{fprac} for \gls{rlnc}.

\subsection{Recovery Process}
\label{sec:snc-recovery-process}
The main difference in the recovery process of \gls{snc} is the method of finding dependent packets. 
This is due to the fact, as the sparsity of coefficients increases, the chance of receiving more dependent packets increases as well~\cite{8631177}.  
 
 
First, a reception matrix denoted by $M'$ is formed with packets in $B_{invalid}$ and $B_{valid}$. 
Each row of $M'$ contains $g$ coefficient elements followed by $l$ coded symbols of a coded packet. 
First, we perform Gaussian elimination over $M'$.
If $M'$ contains any dependent group, there must be a row in which the first $g$ elements of the row are zero.
A vector is devoted to every packet in $M'$.
$V_{i}$ refers to the associated vector with the packet in the $i$th row of the initial form of $M'$. 
During the Gaussian elimination, we keep the index of packets involved in the row elementary operation of other packets in their associated vector.
Upon the Gaussian elimination process, if there is any row where its first $g$ elements are zero, its associated vector indicates a group of dependent packets.
Similar to the Fly-\gls{prac} recovery process for \gls{rlnc}, the last $l$ element of the row indicates the estimation of error locations for the group of dependent packets.
The pseudocode of finding a dependent group is provided in \autoref{alg:dependent-group}.

	


\begin{algorithm}
    \DontPrintSemicolon
    \SetKwInOut{Input}{Input}
    \Input{\;
    $r_{i}$ : $i$th row of $M'$\;
    $a_{i,j}$ : the element in the $i$th row and $j$th column\;
    $m$ : number of row in $M'$\;
    $P_{r}=1$ : pivot  row initialization\;
    $P_{k}=1$ : pivot  column initialization\;
    index($r_{i}$) : gives the index of packet in the $i$th row\;
    $V_{i}$ : refers to the associated vector with the packet whose index is $i$\;}
    \BlankLine
    \Begin{    
    \While{$P_{r}<m$ and $P_{k}<g$}{
        $I \gets argmax_{i}(i = P_{r}~\textit{to}~m, a_{i,P_{k}})$\;
        \If{$a_{I,P_{k}} = 0$}{
            $P_{k} = P_{k} + 1$.
        }\Else{
        swap($r_{P_{r}}$,$r_{I}$)\;
        \For{$i=1$~to~$m-1$}{
            \For{ $j=i+1$~to~$m$}{
                scalar $\gets a_{j,i}/a_{i,i}$\;
                \If{$scalar \neq 0$}{
                    $r_{j} \gets  r_{j} - (scalar \times r_{i})$\;
                    $\theta \gets \textit{index}(r_{i})$\;
                    $\gamma \gets  \textit{index}(r_{j})$\;
                    $V_{\theta} \gets  V_{\theta} \bigcup V_{\gamma}$\;
                    } 
                }
            }
        }
        $P_{k} = P_{k} + 1$\;
        $P_{r} = P_{r} + 1$\;
    }
    ANS $\gets$ return a row where its first $g$ elements are zero\;
    \If{ANS $\neq \emptyset$ }{
        $I \gets$ index(ANS)\;
        $D \gets V_{I} \bigcup \{I\} $ :  $D$ is a set of indices of packets in the dependent group\;
        $L \gets$ last $l$ elements of ANS : L is a set containing estimation of the corrupted locations for packets in $D$\;
        return $(D,L)$\;
        }
    }
    \caption{Finding Dependent Group\label{alg:dependent-group}}
\end{algorithm}

The recovery process is similar to the recovery process of \gls{fprac} for \gls{rlnc}:
This process begins if there are two or more invalid packets among packets involved in the reported dependent group. 
Once $g$ number of linearly independent valid or recovered packets are in $B_{valid}$, the receiver terminates the recovery process and decodes all coded packets using an \gls{snc} decoder.

\section{Probability of Failure in Estimation and Recovery Processes}
\label{sec:parameter-analysis}

In this section, we analyze the impact of the different parameters on the performance of recoverability of the present work. 
Network coding packet recovery methods cannot correct all received erroneous packets. This can be due to: 
\begin{itemize}
    \item Failure to identify all erroneous bits of packets
    \item Falsely verifying the correctness of an erroneous packet
\end{itemize}

We refer to these two events as false positive events in the estimation and correction procedure. In general, we define the recoverability ratio as the number of recovered packets divided by the total received corrupted packets.
The aforementioned false-positive events can affect the recoverability ratio in Fly-PRAC.
According to the proposed method, for a successful recovery, both the estimation and correction processes must be carried out without false-positive events.
If this is not the case, some of the corrupted packets can not be recovered.
In the rest of this section, the probability of such events is calculated, and a general guide for tuning parameters is provided.

\subsection{False-Positive Events in Estimation Process}
\label{sec:false-positive-events-ep}
The estimation process for a group of packets has $l$ rounds, where $l$ is the number of coded symbols in a packet.
For instance, the $i$th round of estimation checks the validity of a column of symbols containing all $i$th coded symbols.
We refer to an erroneous column as an inconsistent column.
During the estimation process, \gls{fprac} may falsely verify the validity of an inconsistent column.
This is defined as a false-positive event during the estimation process.
In this subsection, we investigate the impact of the size of dependent groups~($R$) on the expected inconsistent columns and the probability of false-positive events in the estimation process of \gls{fprac}.

The probability of receiving an error-free symbol can be calculated by \autoref{eq:phs}, and is denoted by $P_{h}^{s}$
\begin{equation}
    P^{s}_{h} = (1- \epsilon ) ^ {b},
    \label{eq:phs}
\end{equation}
where $b$ refers to the bit size of each symbol.

The probability of receiving an error-free column in a dependent group with $R$ packets can be computed by \autoref{eq:phc} and is denoted by $P_{h}^{c}(R)$.
\begin{equation}
    P^{c}_{h}(R) = (P^{s}_{h}) ^ {R}.
    \label{eq:phc}
\end{equation}
Here, we compare the expected number of inconsistent columns in \gls{fprac} with methods based on \gls{acr}.
According to~\cite{PRAC}, $g+1$ packets are required in each round of \gls{acr}, and the probability of an arbitrary column to be error-free is equal to $(P^{s}_{h}) ^{g+1}$.

\begin{figure}[t]
    \vspace*{3 ex}
    \begin{center}
        \begin{tikzpicture}[scale=1]
            \begin{axis}[
                xmode=log,
                height=3.5cm,
                width=8cm,
                legend style={font=\fontsize{7}{5}\selectfont,
                              at={(0.82,0.52)},
                              anchor=north,
                              legend cell align={left}
                              },
                xlabel=$R$,
                ylabel=$ E_{\textit{ic}}$]
                \addplot[mark=triangle*,tudred100] plot coordinates {
                    (11,4.213)
                    (21,7.735)
                    (51,16.757)
                    (101,16.7575)
                };
            \addlegendentry{ACR}
                \addplot[dashed,thick,tudblue100] plot coordinates {
                    (11,16.7575)
                    (21,16.7575)
                    (51,16.7575)
                    (101,16.7575)
                };
                \addlegendentry{Fly-PRAC}
            \end{axis}
        \end{tikzpicture}
    \end{center}
    \caption{The comparison of the expected number of inconsistent columns for different $R$ (colored) with \gls{acr} (red) in a channel with $\epsilon$ =$ 10^{-3}$, where $g=100$, $l=50$, and field size is $GF(2^{8})$. }
    \label{fig:EIC}
\end{figure}
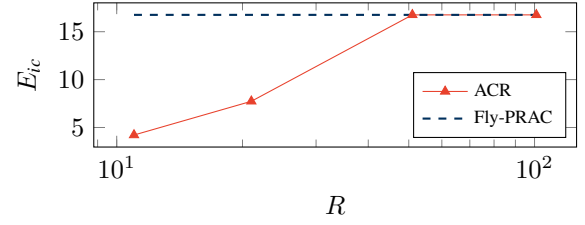

While in \gls{fprac}, in each round of the estimation process, $R$~packets are involved, where $2<R \leq g+1$.
As we decrease~$R$, the probability of receiving an error-free column rises.
Further, the expected number of inconsistent columns ($E_{ic}$) for a group of $R$ packets, each containing $l$ symbols and $s$ segments, is calculated by \autoref{eq:eic}.
\begin{equation}
    E_{\textit{ic}}= \sum_{i = 1}^{l} {l \choose i} \times i \times (P^{c}_{h})^{l - i}  (1 - P^{c}_{h})^{i}
    \label{eq:eic}
\end{equation}
\autoref{fig:EIC} shows a comparison of $E_{ic}$ between proposed estimation process with different~$R$ and \gls{acr}.
	
A fall in $P^{c}_{h}(R)$ negatively affects the recovery process by raising $E_{ic}$, and consequently raising the search space.
The recovery process performs over each invalid segment at a time by permuting the value of symbols in inconsistent columns.
There are $2^{b}$ permutations for each symbol. 
The expected number of inconsistent columns for each segment is equal to $\frac{E_{ic}}{s}$.
The maximum number of permutations for an invalid segment with $\frac{E_{ic}}{s}$ symbols in inconsistent columns is equal to $(2^{b})^{\frac{E_{ic}}{s}}$. 

In some cases, the estimation process falsely verifies an inconsistent column.
In the following, we calculate the probability of such events.
	
Suppose each symbol has eight bits.
Also, assume $e_{1},e_{2},\ldots, e_{R}$ refers to error patterns of symbols involved in each round. 
For instance, $e_{i}= \{01001000\}$ describes that in the $i$th symbol of this round, the second and fifth bits are corrupt.
Let $P_{even}$ referring to the probability that there are an even number of one element among $i$th bits of error patterns.
$P_{even}$ can be calculated according to \autoref{eq:peven}.
\begin{equation}
    P_{even} = \sum_{i = 1}^{\floor{\frac{R}{2}}} {R \choose 2i} (1-\epsilon)^{R-2i} (\epsilon)^{2i}
    \label{eq:peven}
\end{equation}

Moreover, let $P_{0}$ refer to the probability that all $i$th bits of error patterns are zero.
$P_{0}$ can then be calculated by \autoref{eq:p0}. 
\begin{equation}
    P_{0} = (1-\epsilon)^{R}
    \label{eq:p0}
\end{equation}

Regarding the encoding process of a dependent group, a false-positive event happens if $e_{1} \oplus e_{2} \oplus \cdots \oplus e_{R} = \{00000000\}$, and there exists $e_{i}$ where $e_{i} $ contains at least one erroneous bit.
Therefore, the probability of a false-positive event in a round of the estimation process is calculated as follows with \autoref{eq:pfpe}.
\begin{equation}
    P^{(1)}_{fpe} = \sum_{i = 1}^{b} {b \choose i} (P_{even})^{i} (P_{0})^{b-i}
    \label{eq:pfpe}
\end{equation}
	
\begin{figure}[!t]
	\vspace*{3ex}
	\vspace{10px}
		\begin{center}
			\begin{tikzpicture}[scale=1]
				\begin{axis}[
					ymode=log,
					xmode =log,
					height=6cm,
					width=8cm,
					legend style={font=\fontsize{6}{5}\selectfont,
                                  at={(0.77,0.39)},
                                  anchor=north,
                                  legend cell align={left}
                                  },
					xlabel=$\epsilon$,
					ylabel=$ P^{(1)}_{fpe}$]
					\addplot[mark=square*,tudred100] plot coordinates {
						(0.0001,1.9e-06)
						(0.001,1.8e-04)
						(0.01,1.5e-02)
					};
					\addlegendentry{Simulation - GF($2$)}

					\addplot[mark=triangle*,tudgreen100] plot coordinates {
						(0.0001,1.8e-06)
						(0.001,1.8e-04)
						(0.01,1.5e-02)
					};
					\addlegendentry{Formula - GF(2)}

					\addplot[mark=square*,tudblue100] plot coordinates {
							(0.0001,1.4e-05)
							(0.001,1.3e-03)
							(0.01,3.3e-02)
					};
					\addlegendentry{Simulation - GF($2^8$)}

					\addplot[mark=triangle*,tudorange100] plot coordinates {
						(0.0001,1.4e-05)
						(0.001,1.2e-03)
						(0.01,3.3e-02)
					};
					\addlegendentry{Formula- - GF($2^8$)}
				\end{axis}
			\end{tikzpicture}
		\end{center}
	\caption{The figure shows the comparison of the probability of false-positive events resulting from \autoref{eq:pfpe} and from simulation results for different $\epsilon$ and $R=20$.}
	\label{fig:estimationfp}
\end{figure}
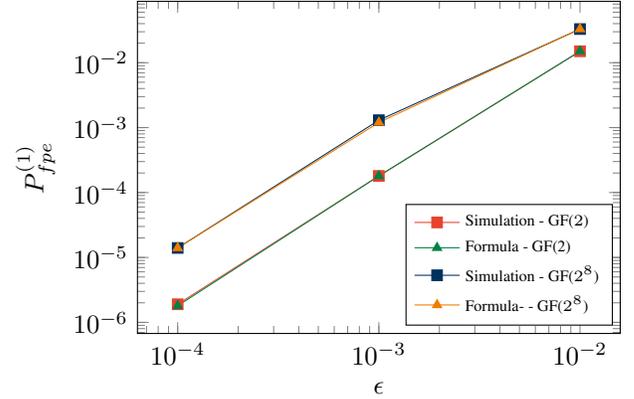

\autoref{fig:estimationfp} compares the outcome of \autoref{eq:pfpe} with simulation results for the finite fields GF($2$) and GF($2^{8}$).
\begin{figure}[t]
    \vspace*{3 ex}
    \begin{center}
        \begin{tikzpicture}[scale=1]
            \begin{axis}[
                legend columns=-1,
                xmode=normal,
                ymode=log,
                height=3.5cm,
                width=8cm,
                legend style={font=\fontsize{7}{5}\selectfont,
                              at={(0.5,1.4)},
                              anchor=north,
                              legend cell align={left}
                              },
                xlabel=$R$,
                ylabel=$ P_{fpe}^{(1)}$]
                
                \addplot[mark=triangle*,tudred100] plot coordinates {
                    (5,7.996960590722312e-09)
                    (10,3.5971931391418406e-08)
                    (20,1.517600389068457e-07)
                    (30,3.471727957381162e-07)
                    (40, 6.220189989396262e-07)
                    (50,9.761077689499657e-07)
                };
                \addlegendentry{$\epsilon = 10^{-5}$}

                \addplot[mark=triangle*,thick,tudblue100] plot coordinates {
                    (5,7.969659002365473e-07)
                    (10,3.5720336284419228e-06)
                    (20,1.496181882228002e-05)
                    (30,3.398203677653302e-05)
                    (40,6.044852417089019e-05)
                    (50,9.418022677501002e-05)
                };
                \addlegendentry{$\epsilon = 10^{-4}$}

                \addplot[mark=triangle*,thick,tudorange100] plot coordinates {
                    (5,7.701831057118731e-05)
                    (10,0.00033302830934941807)
                    (20,0.0012986473383958402)
                    (30,0.002746980070206817)
                    (40,0.004552492107607566)
                    (50,0.0066105697022532375)
                };
            \addlegendentry{$\epsilon = 10^{-3}$}
            \end{axis}
        \end{tikzpicture}
    \end{center}
    \caption{The figure shows the probability of false-positive events during the estimation process for an arbitrary column of $R$ packets for different bit error rates.}
    \label{fig:ROnFPE}
\end{figure}
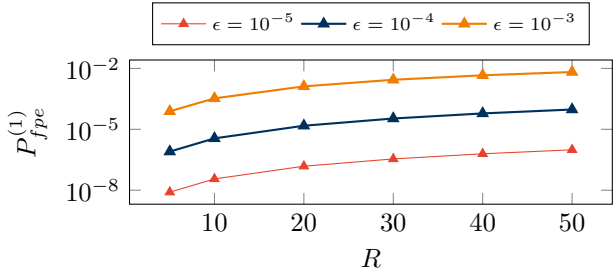

\autoref{fig:ROnFPE} shows the probability of a false positive event for an arbitrary column during the estimation problem according to the formula \ref{eq:pfpe}.
The \autoref{fig:ROnFPE} depicts that as the number of $R$ increases, $P^{(1)}_{fpe}$ increases as well.
This shows, for a channel with higher error rates, the $R$ size should be reduced in order to reduce false-positive events during the estimation process.

Any failure in one of the estimation rounds results in the failure of the recovery for at least one packet.
In other words, since the location of the error is not detected, no correction process is carried out for that section and thus packets with undetected error locations cannot be fixed.
Since the estimation procedure is carried out individually for $l$ different groups of symbols, the chance of no estimation failure is equal to no estimate false-positive for $l$ estimation rounds.
This can be calculated per \autoref{eq:failureestimation}.
\begin{equation}
    P_{fe} = (1- P^{(1)}_{fpe})^{l}
    \label{eq:failureestimation}
\end{equation}

In the following, we demonstrate that, for a constant bit error rate, decreasing $R$,  reduces the probability of a false-positive during the estimation process increases as well.
In addition, section VII provides more simulations on the effect of $R$ on the performance of \gls{fprac} and the packet recoverability ratio.

\subsection{False-Positive Events in Correction Process}
\label{sec:false-positive-events-cp}
Although, \gls{crc} is a compelling solution for error detection; \gls{crc} cannot detect all errors.
Since the correction process uses \gls{crc} checks to verify the integrity of a segment, the \gls{crc} check might falsely verify erroneous data.
We refer to it as a false-positive event in the correction process, and this subsection discusses the probability of such events.

\gls{fprac} does the correction process for segments that \gls{icrc} refutes their validity.
However, \gls{icrc} may not detect an erroneous segment.
Let $N_{p}$ refer to the number of bits of a segment and its associated \gls{icrc}.
According to a given bound for the probability of undetected errors of a \gls{crc} in~\cite{CRCDetection}, the probability that an \gls{icrc} does not identify an erroneous segment can be calculated using \autoref{eq:undetectedCRC}

\begin{equation}
	P_{un} = \sum_{i= 1}^{N_{p}} A^{i}_{w} (1-\epsilon)^{N_{p}-i} (\epsilon)^{i},
	\label{eq:undetectedCRC}
\end{equation}
where $A^{i}_{w}$ refers to the number of valid code words with the Hamming weight of $w$.
\autoref{eq:undetectedCRC} clearly shows the impact of \gls{crc} type and data length on $P_{un}$.

Also, in the correction, \gls{fprac} changes the value of symbols reported by the estimation process, and after each change, it checks its integrity with associated \gls{icrc}.
In the following, we investigate the probability of false-positive events during these correction trials.
Let $V_{p}$ and $V'_{p}$ represent vectors of size $N_{p}$, respectively, holding the error-free and received versions of segment data in binary.
Also, let $E$ be a vector of size $N_{p}$ representing error locations with a ``1'' element, while other error-free locations are zeros.
\begin{equation}
	V_{p} \oplus E = V'_{p}.
\end{equation}

Upon receiving a packet, the integrity of information is checked by \gls{icrc}, and if it is inconsistent with the data, the recovery process begins.
In correction trials, the value of $V'_{p}$ is changed until the \gls{icrc} verifies its integrity. 
For this reason, an XOR operation is performed on $V'_{p}$ and $X$, where $X$ is a vector of the same size as $V'_{p}$ containing zero and one elements.
After each change, the associated \gls{icrc} checks the integrity of the resulting vector $V''_{p}$.
\begin{equation} 
	V'_{p} \oplus X = V''_{p}.
\end{equation}
Afterwards, if the \gls{icrc} verifies the resulting vector ($V''_{p}$) as an error-free version of $V'_{p}$, the correction ends, and $V''_{p}$ is replaced with $V'_{p}$ in the packet.
On the other hand, if the \gls{icrc} refutes the integrity of $V''_{p}$, this process repeats with different $X$.

Assume we are using a \gls{crc} with linearity properties, i.e., the XOR of two valid code words yields another valid code word.
If $E$ is not a zero vector and $X \oplus E$ constitutes a valid code word, the \gls{icrc} verifies it.
Since $V''_{p} \neq V_{p} $, the \gls{icrc} has verified erroneous data, which we refer to it as a false-positive event.

Let $\alpha_{1}$ and $\alpha_{2}$ refer to the Hamming weight of $E$ and $X$ in turn.
Clearly, the Hamming weight of $E \oplus X$ is ranging from $|\alpha_{1} - \alpha_{2}|$ to $\min(N_{p},\alpha_{1} + \alpha_{2})$.
Suppose $h_{d}$ to be the Hamming weight of $E \oplus X$, therefore the probability of $E \oplus X$ becomes a valid code word which is denoted by $P_{valid}$ and can be calculated by \autoref{eq:tovalid}.
\begin{equation}
	P_{valid} = \frac{A_{w}^{h_{d}}}{{N_{p} \choose h_{d}}}.
	\label{eq:tovalid}
\end{equation}

If an $X$ with the weight of $\alpha_{2}$ XOR with $V'_{p}$ containing $\alpha_{1}$ bit errors and suppose $j$ is the number of elements in $X$ that are ``1'' at elements where their corresponding elements in $V'_{p}$ are erroneous, then as a result, the Hamming distance between $V''_{p}$ and $V_{p}$ would be $\alpha_{1} + \alpha_{2} - 2j$. 
The value of $j$ and $\alpha_{1} +\alpha_{2}-2j $ can not be more than $N_{p}$.
The minimum value of $j$ can be obtained using \autoref{eq:j}, and it is denoted by $j_{\min}$.
\begin{equation} 
    j_{\min} = 
    \begin{cases} 
    	0 & \alpha_{1} + \alpha_{2} - N_{p} \leq 0\\
    	\alpha_{1} + \alpha_{2} - N_{p} & \text{otherwise}.\\
	\end{cases}
	\label{eq:j}
\end{equation}
The recovery algorithm starts by using permutations of $X$ with minimum weight, and obviously, the correct answer is found where $\alpha_{1} = \alpha_{2}$.
Thus, $\alpha_{2}$ is increased from 1 to $\alpha_{1}$.
Also, there are ${\alpha_{2} \choose j}{N_{p} - \alpha_{2}  \choose  \alpha_{2} - j}$ possible correction trials for an $X$ of weight $\alpha_{2}$. 

Let $P_{nq}$ indicate the probability of no false-positive events during all correction trials using $X$ of weight $ \alpha_{2} $, where $ \alpha_{1} \neq \alpha_{2}$.
Then, it can be calculated using the \autoref{eq:1}.
\begin{equation}
	P_{nq}=1 - \prod_{j=j_{min}}^{\alpha_{2}}
	\left(1- {\frac{A_{w}^{\alpha_{1} + \alpha_{2}-2j}}
		{{N_{p} \choose \alpha_{1} + \alpha_{2}-2j}}}
	\right)^{{\alpha_{2} \choose j}{N_{p} - \alpha_{2} \choose \alpha_{2}-j}}.
	\label{eq:1}
\end{equation}

Furthermore, let $P_{eq}$ indicate the probability of no false-positive events, when during all correction trials using $X$ of weight $ \alpha_{2} $, where $ \alpha_{1} =\alpha_{2}$.
The value of $P_{eq}$ can be calculated using the \autoref{eq:2}.
\begin{equation}
	P_{eq}=1 - \prod_{j=j_{min}}^{\alpha_{2}-1}
	\left(1- {\frac{A_{w}^{\alpha_{1} + \alpha_{2}-2j}}
		{{N_{p} \choose \alpha_{1} + \alpha_{2}-2j}}}
	\right)^{{\alpha_{2} \choose j}{N_{p} - \alpha_{2} \choose \alpha_{2}-j}}.
	\label{eq:2}
\end{equation}

In all, the probability of no false-positive events, where $E$ has a Hamming weight of $\alpha_{1}$, can be calculated using \autoref{eq:3}, and is denoted by $P_{nfp}$.
\begin{equation}
	P_{nfp} = \prod_{\alpha_{2}=0}^{\alpha_{1}} f(\alpha_{1},\alpha_{2}),
	\label{eq:3}
\end{equation}
where
\[ f(\alpha_{1},\alpha_{2}) = 
\begin{cases} 
	P_{nq} & \alpha_{1} < \alpha_{2} \\
	P_{eq} & \alpha_{1} = \alpha_{2} \neq 0 \\
	1 & \alpha_{1} = \alpha_{2} = 0.\\
\end{cases}
\]

In addition, the probability of no false-positive events or successful recovery of a segment is denoted by $P_{\textit{sr}}$, and is calculated per \autoref{eq:4}.
\autoref{fig:NofpC} compares the outcome of \autoref{eq:4} with simulation results.
\begin{equation}
	P_{sr}= \sum_{\alpha_{1} = 0}^{N_{p}} (1-\epsilon)^{N_{p}-\alpha_{1}} (\epsilon)^{\alpha_{1}} (\prod_{\alpha_{2}=0}^{\alpha_{1}} f(\alpha_{1},\alpha_{2})).
	\label{eq:4}
\end{equation}

\begin{figure*}
    \begin{minipage}[t]{0.49\textwidth}
        \centering
        \begin{tikzpicture}[scale=1]
            \begin{axis}[
                xmode=log,
                height=4.5cm,
                width=8cm,
                legend style={font=\fontsize{6}{5}\selectfont,
                              at={(0.24,0.54)},
                              anchor=north,
                              legend cell align={left}
                              },
                xlabel=$\epsilon$,
                ylabel=$ P_{sr}$]
                \addplot[mark=square*,tudred100] plot coordinates {
                    (0.000001,0.999986)
                    (0.00001, 0.999860)
                    (0.0001, 0.998600)
                    (0.001, 0.986091)
                };
                \addlegendentry{ Formula 17 - $N_{p}$  = 15}
                
                \addplot[mark=triangle*,tudgreen100] plot coordinates {
                    (0.000001,0.999986)
                    (0.00001,0.99994)
                    (0.0001,0.999254)
                    (0.001,0.989072)
                };
                \addlegendentry{Simulation - $N_{p}$  = 15}
                
                \addplot[mark=square*,tudblue100] plot coordinates {
                    (0.000001,0.999978)
                    (0.00001,0.9997800)
                    (0.0001, 0.997802)
                    (0.001,0.978230)
                };
                \addlegendentry{ Formula 17  - $N_{p}$  = 23}
                
                \addplot[mark=triangle*,tudorange100] plot coordinates {
                    (0.000001,0.999968)
                    (0.00001,0.999714)
                    (0.0001,0.997564)
                    (0.001,0.97619)
                };
                \addlegendentry{Simulation - $N_{p}$  = 23}
            \end{axis}
        \end{tikzpicture}
        \caption{The figure depicts the probability of a successful recovery of a segment ($P_{sr}$) at different $\epsilon$, where the \gls{crc}'s divisor is \texttt{0xe7}.}
	    \label{fig:NofpC}
    \end{minipage}
    \hfill
    \begin{minipage}[b]{0.49\textwidth}
        \centering
        \renewcommand{\arraystretch}{1}
    	\captionof{table}{The table presents the goodput [\si{\kilo\bit\per\second}] comparison of \gls{fprac}, \gls{sprac}, and \gls{rlnc} without partial packet recovery where the payload size is \SI{900}{\byte}, the data rate is \SI{500}{\kilo\bit\per\second}, and $g=50$.}
    	\label{table:BERvsGoodoput}
        \resizebox{\columnwidth}{!}{%
        \begin{tabular}{|c|c|c|cccc|}
            \hline
            \multirow{2}{*}{Method}   & \multirow{2}{*}{$R$} & \multirow{2}{*}{$s$} & \multicolumn{4}{c|}{$\epsilon$}                                                                     \\ \cline{4-7} 
            &                    &                    & \multicolumn{1}{c|}{ $10^{-5}$ }   & \multicolumn{1}{c|}{$5\times 10^{-5}$}  & \multicolumn{1}{c|}{ $10^{-4}$}   &  $5 \times 10^{-4}$\\ \hline
            \multirow{2}{*}{\acrshort{fprac}} & 25                 & 10                 & \multicolumn{1}{c|}{427.5} & \multicolumn{1}{c|}{418.7} & \multicolumn{1}{c|}{406.2} & 134.1 \\ \cline{2-7} 
            & 10                 & 30                 & \multicolumn{1}{c|}{401}   & \multicolumn{1}{c|}{390}   & \multicolumn{1}{c|}{388.3} & 345.3 \\ \hline
            \acrshort{sprac}                    & -                  & 10                 & \multicolumn{1}{c|}{447}   & \multicolumn{1}{c|}{396}   & \multicolumn{1}{c|}{198}   & 0     \\ \hline
            \acrshort{rlnc}                      & -                  & -                  & \multicolumn{1}{c|}{460}   & \multicolumn{1}{c|}{348}   & \multicolumn{1}{c|}{234.7} & 12.5  \\ \hline
        \end{tabular}%
        }
    \end{minipage}

    \begin{minipage}{\textwidth}
        \vspace{1cm}
    \end{minipage}

    \begin{minipage}[t]{0.49\textwidth}
        \centering
        \begin{tikzpicture}[scale=1]
			\begin{axis}[
				height=3.5cm,
				width=8cm,
				legend style={font=\fontsize{7}{5}\selectfont,
                              at={(0.18,0.95)},
                              anchor=north,
                              legend cell align={left}
                              },
				xlabel={$Generation~Size~(g)$},
				ylabel={$\acrshort{ct}~[ms]$}
                ]
				\addplot[mark=triangle*,tudblue100] plot coordinates {
					(50,303)
					(100,903.7)
					(150,2319.3)
					(200,4897.17)
				};
				\addlegendentry{\acrshort{fprac}}
				
				\addplot[color=tudred100,mark=*]
				plot coordinates  {
					(50,1438.9)
					(100,2368.2)
					(150,4308.9)
					(200,6192.31)
				};
				\addlegendentry{\gls{sprac}}
			\end{axis}
		\end{tikzpicture}
		\caption{The figure shows the \gls{ct} comparison of \gls{fprac} with $R=10$, and \gls{sprac}, where $\epsilon = 5 \times 10^{-5}$, data rate = \SI{5}{\mega\bit\per\second}, $s = 4$ and payload size is \SI{800}{\byte}.}
        \label{fig:completionG1}
    \end{minipage}
    \hfill
    \begin{minipage}[t]{0.49\textwidth}
        \centering
        \begin{tikzpicture}
    		\begin{axis}[
    			ymode=normal,
    			height=3.5cm,
    			width=8cm,
    			legend style={font=\fontsize{7}{5}\selectfont,
                              at={(0.18,0.95)},
                              anchor=north,
                              legend cell align={left}
                              },
    			xlabel={$Payload~Size~[B]$},
    			ylabel={$\acrshort{ct}~[ms]$}
                ]
                \addplot[mark=triangle*,tudblue100] plot coordinates {
    				(300,667.1)
    				(500,748.6)
    				(700,826.0)
    				(900, 910.3)
    			};
    			\addlegendentry{\acrshort{fprac}}
    			
    			\addplot[color=tudred100,mark=*]
    			plot coordinates  {
    				(300,836.5)
    				(500,1541.7)
    				(700,2168.3)
    				(900,2544.8)
    			};
    			\addlegendentry{\gls{sprac}}
    		\end{axis}
    	\end{tikzpicture}
        \caption{\gls{ct} comparison between \gls{fprac} with $R=10$ and \gls{sprac} where $s=4$ , data rate = \SI{5}{\mega\bit\per\second}, $\epsilon = 5 \times 10^{-5}$ and $g=100$.}
        \label{fig:completionP}
    \end{minipage}
\end{figure*}

Finally, the probability of a false-positive event during correction trials for a segment can be calculated by \autoref{eq:fpcs}.
\begin{equation}
	P_{fpc}^{(1)}= 1 - P_{sr}
	\label{eq:fpcs}
\end{equation}

In summary, for a given \gls{crc} type and bit error rate, as the codeword length, or in this context, the segment size, increases, the probability of undetected error increases as well~\cite{crcratio}.
In addition, we have shown in \autoref{fig:NofpC}, for a given segment size and \gls{crc}, as the error rate increases, the probability of successful recovery decreases, and thus, according to \autoref{eq:fpcs}, the probability of undetected error increases, too.
Therefore, depending on the required level of reliability, channel condition, segment size, or \gls{crc} type should be selected.
Further simulations on the impact of segment size on the packet recovery ratio and \gls{fprac} are discussed in \autoref{sec:simulation-results}.

\subsection{Failure in the Recovery Process}
\label{sec:failure-in-recovery-process}

\section{Simulation and Results}
\label{sec:simulation-results}
In this section, we compare the performance of \gls{fprac} with its most notable related works, i.e., \gls{sprac} and \gls{rlnc} for various settings.
The effect of generation size and payload size, i.e., the total size of symbols in a packet,  on goodput in a point-to-point \gls{rlnc} transmission is reported.
Furthermore, the impact of enabling recoding and recovery for the intermediate node is investigated.
Additionally, the efficiency of packet recovery in terms of goodput and \gls{add}~\cite{amir} for \gls{snc}-based communication is examined.
Finally, these methods are compared in terms of recoverability and overhead.
	
\subsection{Experimental Setup}
\label{sec:experimental-setup}
All methods are implemented using the \textit{Kodo} library~v16.1.1~\cite{Kodo}.
Each simulation is repeated at least 1000 times, and the average is reported.
Experiments were performed on a machine running Ubuntu~20.04~LTS with an Intel Core i7-4710HQ and \SI{8}{\giga\byte} of DDR3 RAM.
We consider a field size of $GF(2^8)$ throughout the experiments.
The reported \acrfull{ct} includes the transmission and decoding process.  

\subsection{Impact of Generation and Payload Size on Packet Recovery}
\label{sec:impact-generation-payload-size-on-pr}
We compare the performance of \gls{fprac} with \gls{rlnc} without partial packet recovery and \gls{sprac} in terms of goodput for different error rates.
The effect of generation and payload sizes on the \gls{ct} of \gls{fprac} and \gls{sprac} is investigated as well.
	 
\autoref{table:BERvsGoodoput} compares the goodput of \gls{fprac} with \gls{sprac} and \gls{rlnc} without \acrfull{ppr}.
\autoref{table:BERvsGoodoput} shows that when $\epsilon$ is as low as $10^{-5}$, \gls{rlnc} without \gls{ppr} has a higher goodput.
However, as the error rate increases, the goodput of \gls{rlnc} reduces significantly.
For instance, when the data rate is \SI{500}{\kilo\bit\per\second}, and the error rate is $10^{-4}$, the goodput of \gls{fprac} with $R=25$ is \SI{406.2}{\kilo\bit\per\second} compared to \SI{198}{\kilo\bit\per\second} and \SI{234.7}{\kilo\bit\per\second} for \gls{sprac} and \gls{rlnc}, respectively.
	

\autoref{fig:completionG1} illustrates that for larger generation sizes, it takes more time to decode a generation due to the higher complexity of decoding.
It can be observed, that for \gls{fprac} with $R=10$, the \glspl{ct} are \SI{303}{\milli\second} and \SI{903.7}{\milli\second}, when the generation sizes are $50$ and $100$, respectively.

Generally, larger payload sizes have more inconsistent columns, which makes the recovery process more time-consuming.
Moreover, \autoref{fig:completionP} shows, the increase is more dramatic for \gls{sprac}.
For instance, increasing the payload size from $300$ to \SI{900}{\byte} leads to an increase in \gls{ct} of \gls{sprac} from \SI{836.5}{\milli\second} to \SI{2544.8}{\milli\second}.
While, for \gls{fprac} with $R=10$, the \glspl{ct} for the same payload sizes are increased from \SI{667.1}{\milli\second} to \SI{910.3}{\milli\second}.
	
\subsection{Impact of $R$ and Segment Size on Packet Recovery}
\label{sec:impact-r-number-segments-pr}
We investigate the impact of the number of segments and $R$ on goodput, and the percentage of recovered packets.

\autoref{fig:segVSgoodput} shows that increasing the number of segments will boost goodput up to a certain point; however, adding too many segments to each packet is unnecessary, since it can degrade goodput due to its overhead.
For example, by increasing the number of segments from $2$ to $25$, the goodput increases from \SI{144.8}{\kilo\bit\per\second} to \SI{214.2}{\kilo\bit\per\second}.
Furthermore, a growth in the number of segments increases the chance of recovery.
\autoref{fig:segVSrecoverability} shows that the percentage of recovered packets for \gls{fprac} with $2$ and $25$ segments are $29\%$ and $96\%$, respectively.

\begin{figure*}
    \begin{minipage}[t]{0.49\textwidth}
        \centering
        \begin{tikzpicture}
            \begin{axis} [ 
                bar width=10pt,
                ymin=130,ymax=225,
                height=4cm,
                width=8cm,
                xtick=data,
                x tick label as interval=false,
                ylabel={$Goodput~[\unit{\kilo\bit\per\second}]$},
                xlabel={$Number~of~Segments~s$}
                ]
                \addplot[tudblue100,mark=*] coordinates {
                    (2,146.8) 
                    (5,207.2) 
                    (10,216.1) 
                    (25,217.1) 
                    (50,212.0) 
                    (75,207.4)
                    (100,201.7)  
                };
            \end{axis}
        \end{tikzpicture}
        \caption{In this figure, the goodput comparison of \gls{fprac} for different number of segments is shown, where $g=50$, payload size is \SI{900}{\byte}, the data rate is \SI{250}{\kilo\bit\per\second}, $\epsilon = 10^{-4}$, and $R = 25$.}
    \label{fig:segVSgoodput}
    \end{minipage}
    \hfill
    \begin{minipage}[t]{0.49\textwidth}
        \centering
        \begin{tikzpicture}
            \begin{axis} [ 
                bar width=10pt,
                ymin=20,ymax=105,
                height=4cm,
                width=8cm,
                xtick=data,
                x tick label as interval=false,
                ylabel={$Recovery~Ratio~[\%]$},
                xlabel={$Number~of~Segments~s$}
                ]
                \addplot[tudblue100,mark=*] coordinates {
                    (2,29) 
                    (5,81) 
                    (10,90) 
                    (25,96) 
                    (50,97) 
                    (75,98)
                    (100,98)
                };
            \end{axis}
        \end{tikzpicture}
        \caption{The figure presents the impact of number of segments on the percentage of recovered packets where $g=50$, payload size is \SI{900}{\byte}, the data rate is \SI{250}{\kilo\bit\per\second}, $\epsilon= 10^{-4}$, and $R = 25$.}
        \label{fig:segVSrecoverability}
    \end{minipage}

    \begin{minipage}[t]{0.49\textwidth}
        \centering
        \begin{tikzpicture}
            \begin{axis} [
                bar width=10pt,
                ymin=155,ymax=200,
                height=4cm,
                width=8cm,
                xtick=data,
                x tick label as interval=false,
                ylabel={$Goodput~\unit{\kilo\bit\per\second}$},
                xlabel={$R$}
                ]
                \addplot[tudblue100,mark=*] coordinates {
                    (5,167.4) 
                    (10,185.6) 
                    (15,188.8) 
                    (20,189.4) 
                    (25,187.7) 
                    (50,173.3)
                };
            \end{axis}
        \end{tikzpicture}
        \caption{The figure shows the goodput comparison of \gls{fprac} for different $R$ sizes where $g=100$, payload size is \SI{900}{\byte}, the data rate is \SI{250}{\kilo\bit\per\second}, $\epsilon = 10^{-4}$, and $s = 5$.}
        \label{fig:RVSgoodput}
    \end{minipage}
    \hfill
    \begin{minipage}[t]{0.49\textwidth}
        \centering
        \begin{tikzpicture}
            \begin{axis} [
                bar width=10pt,
                ymin=70,ymax=100,
                height=4cm,
                width=8cm,
                xtick=data,
                x tick label as interval=false,
                ylabel={$Recovery~Ratio~[\%]$},
                xlabel={$R$}
                ]
                \addplot[tudblue100,mark=*] coordinates {
                    (5,88.4) 
                    (10,89.6) 
                    (15,86.7) 
                    (20,84.3) 
                    (25,81.9) 
                    (50,72.9)
                };
            \end{axis}
        \end{tikzpicture}
        \caption{The impact of $R$ on the percentage of recovered packets where $g=100$, payload size is \SI{900}{\byte}, the data rate is \SI{250}{\kilo\bit\per\second}, $\epsilon= 10^{-4}$, and $s = 5$.}
        \label{fig:RVSrecoverability}
    \end{minipage}
\end{figure*}
	
\autoref{fig:RVSgoodput} displays that reducing $R$ will enhance communication in terms of goodput up to a certain level.
By lowering $R$ from $50$ to $20$, the goodput grows from \SI{162.4}{\kilo\bit\per\second} to \SI{184.1}{\kilo\bit\per\second}.
However, transmitting too many dependent packets degrades goodput.
\autoref{fig:RVSgoodput} shows that the goodput of \gls{fprac} with $R=25$ and $R=5$ are \SI{184.1}{\kilo\bit\per\second} and \SI{161.6}{\kilo\bit\per\second}, respectively. 
	
Moreover, increasing $R$ raises the chance of recovery.
For example, \autoref{fig:RVSrecoverability} reveals that the percentages of recovered packets for \gls{fprac} with $R=50$ and $R=10$ are $72\%$ and $89\%$, respectively.

\subsection{Impact of Enabling Recovery at Intermediate Nodes}
\label{sec:impact-enabling-recovery-intermediate-nodes}
\renewcommand{\arraystretch}{1.1}
\begin{table}[t!]
	\caption{The table shows the comparison of \gls{fprac} and \gls{sprac}, in terms of \acrfull{ct} and total transmission.
    Both methods are using four segments and communicate through an intermediate node where the payload size is \SI{500}{\byte}.}
	\label{table:relay}
    \begin{center}
    	\begin{tabular}{|c|c|c|c|c|c|}
    		\hline
    		&                          & \multicolumn{2}{c|}{$\epsilon$~=~ $5 \times 10^{-5}$.} & \multicolumn{2}{c|}{$\epsilon$~=~$10^{-4}$} \\ \cline{3-6} 
    		\multirow{-2}{*}{Method}   & \multirow{-2}{*}{R}      & Total Sent            & \acrshort{ct}~[ms]        & Total Sent           & \acrshort{ct}~[ms]        \\ \hline
    		& 5                        & 141.3        & 1126.2        & 181.9       & 1502.5       \\ \cline{2-6} 
    		\multirow{-2}{*}{\acrshort{fprac}} & 10                       & 136.6        & 1221.9        & 202         & 2344         \\ \hline
    		\acrshort{sprac}                     & { -} & 146.1        & 1476.45       & 212.3       & 3069.97      \\ \hline
    	\end{tabular}
    \end{center}
\end{table}

In this subsection, we compare the proposed method \gls{fprac} and \gls{sprac} in a unicast scenario, where the source and destination communicate through an intermediate node.
For the \gls{fprac} scenario, the intermediate node can recode and recover partial packets, while it only works in the store-and-forward mode in the \gls{sprac} scenario.
	
\autoref{table:relay} clearly illustrates that in a poor channel condition, the proposed method \gls{fprac} reduces total transmissions.
In particular, for $\epsilon = 5 \times 10^{-5}$, \gls{fprac} with $R=10$ requires $136.6$ packets on average to decode a generation compared to $146.1$ packets for \gls{sprac}.
	
In \autoref{table:relay}, the time spent by relay, recovering, and recoding packets are considered in the \gls{ct} of \gls{fprac}.
It can be observed that for poor channel conditions, like $\epsilon =10^{-4}$, the \acrfull{ct} of the proposed method \gls{fprac} with $R = 5$ and \gls{sprac} are \SI{1502}{\milli\second} and \SI{3069}{\milli\second}, respectively.
	
\subsection{Comparison of Packet Recovery for SNC Communication}
\label{sec:comparison-packet-recovery-snc-communication}
Here, we evaluate the impact of \gls{fprac} and \gls{sprac} by comparing both methods in terms of \acrfull{add}.
\gls{add} can be calculated by \autoref{eq:add}
\begin{equation}
    ADD = \dfrac{\sum_{i=0}^{g} d_{i}}{g},
    \label{eq:add}
\end{equation}
where $d_{i}$ indicates the number of transmissions until the $i$th packet is decoded.
	
As we decrease the value of $w$, the receiver must collect more packets to obtain $g$ linearly independent packets.
For instance, in \gls{snc} without \acrfull{ppr}, for a channel with $\epsilon=5 \times 10^{-5}$, when $w$ is $2$ and $3$, the average number of required coded packets to fully decode a generation are $372.7$ and $231$, in turn.
	
The proposed method \gls{fprac} exploits redundant packets to begin recovery earlier.
It enhances the \gls{add} of an \gls{snc} scheme in noisy conditions.
\autoref{table:ADD} illustrates that utilizing packet recovery boosts the partial decoding in terms of \gls{add}.
For instance, when $\epsilon=5\times 10^{-5}$, $w=2$, and $g=100$, the \gls{add} in case of discarding partial packets is on average $113.5$ while using \gls{fprac} reduces it to $103.9$.
It also leads to a reduction in the number of required transmissions from $372.7$ to $322.97$ packets.
 
Moreover, results show that \gls{fprac} is more successful in reducing \gls{add} than \gls{sprac}.
For example, \autoref{table:ADD} shows, when $\epsilon = 10^{-4}$, $w=2$ and $g=100$ the \gls{add} of \gls{fprac} and \gls{sprac} are $106.9$ and $156.7$, respectively.
	
\begin{table}[t!]
    \caption{The table shows the comparison of \gls{fprac} (with $R=5$), \gls{sprac}, and \gls{rlnc} in terms of \gls{add} and total transmissions, where $g=100$ and payload size is \SI{800}{\byte}.}
    \label{table:ADD}
    \begin{center}
        \begin{tabular}{|c|c|c|c|c|c|}
            \hline
            $\epsilon$                  & w                  &            & \acrshort{fprac} & \acrshort{sprac} & \acrshort{rlnc} \\ \hline
            \multirow{4}{*}{$5\times10^{-5}$} & \multirow{2}{*}{2} & \acrshort{add}        & 103.9    & 114    & 113.5  \\ \cline{3-6} 
            &                    & Total Sent & 322.97   & 326.7  & 372.7  \\ \cline{2-6} 
            & \multirow{2}{*}{3} & \acrshort{add}        & 118.9    & 129.8  & 131.4  \\ \cline{3-6} 
            &                    & Total Sent & 215.9    & 220.3  & 231    \\ \hline
            \multirow{4}{*}{$10^{-4}$}  & \multirow{2}{*}{2} & \acrshort{add}        & 106.9    & 156.7  & 161    \\ \cline{3-6} 
            &                    & Total Sent & 326.2    & 427    & 485.5  \\ \cline{2-6} 
            & \multirow{2}{*}{3} & \acrshort{add}        & 122.6    & 178    & 180.5  \\ \cline{3-6} 
            &                    & Total Sent & 220.2    & 286.6  & 331.5  \\ \hline
        \end{tabular}
    \end{center}
\end{table}

\subsection{Recoverability and Overhead Comparison}
\label{sec:recoverability-overhead-comparison}
\renewcommand{\arraystretch}{1.1}
\begin{table}[t!]
    \centering
    \caption{The table shows the comparison of \gls{fprac} and \gls{sprac}, with $s = 4$ and $ \epsilon =5 \times 10^{-5}$.}
    \label{table:recoverability}
    \begin{tabular}{|c|c|c|cccc|}
        \hline
        \multirow{3}{*}{Method}   & \multirow{3}{*}{$g$}   & \multirow{3}{*}{$R$}      & \multicolumn{4}{c|}{Payload size [\unit{\byte}]}                                                            \\ \cline{4-7} 
        &                      &                         & \multicolumn{2}{c|}{300}                                    & \multicolumn{2}{c|}{700}               \\ \cline{4-7} 
        &                      &                         & \multicolumn{1}{c|}{Total} & \multicolumn{1}{c|}{Recovered} & \multicolumn{1}{c|}{Total} & Recovered \\ \hline
        \multirow{6}{*}{\acrshort{fprac}} & \multirow{3}{*}{100} & 5                       & \multicolumn{1}{c|}{124.3} & \multicolumn{1}{c|}{5.5}       & \multicolumn{1}{c|}{125}   & 20.4      \\ \cline{3-7} 
        &                      & 10                      & \multicolumn{1}{c|}{111.2} & \multicolumn{1}{c|}{8.5}       & \multicolumn{1}{c|}{112.8} & 22.9      \\ \cline{3-7} 
        &                      & \multicolumn{1}{l|}{15} & \multicolumn{1}{c|}{105.9} & \multicolumn{1}{c|}{10}        & \multicolumn{1}{c|}{111.6} & 19.8      \\ \cline{2-7} 
        & \multirow{3}{*}{200} & 5                       & \multicolumn{1}{c|}{249.4} & \multicolumn{1}{c|}{10.8}      & \multicolumn{1}{c|}{250.5} & 41        \\ \cline{3-7} 
        &                      & 10                      & \multicolumn{1}{c|}{222.4} & \multicolumn{1}{c|}{16.5}      & \multicolumn{1}{c|}{225.5} & 46.3      \\ \cline{3-7} 
        &                      & 15                      & \multicolumn{1}{c|}{211.7} & \multicolumn{1}{c|}{19.7}      & \multicolumn{1}{c|}{223}   & 40        \\ \hline
        \multirow{2}{*}{\acrshort{sprac}}   & 100                  & -                       & \multicolumn{1}{c|}{112.3} & \multicolumn{1}{c|}{0.9}       & \multicolumn{1}{c|}{131.7} & 1.6       \\ \cline{2-7} 
        & 200                  & -                       & \multicolumn{1}{c|}{224.4} & \multicolumn{1}{c|}{1.3}       & \multicolumn{1}{c|}{264}   & 1.5       \\ \hline
        \multirow{2}{*}{\acrshort{rlnc}}     & 100                  & -                       & \multicolumn{1}{c|}{113.2} & \multicolumn{1}{c|}{0}         & \multicolumn{1}{c|}{133}   & 0         \\ \cline{2-7} 
        & 200                  & -                       & \multicolumn{1}{c|}{226.2} & \multicolumn{1}{c|}{0}         & \multicolumn{1}{c|}{266.5} & 0         \\ \hline
    \end{tabular}
\end{table}

\autoref{table:recoverability} gives information about average recovered and transmitted packets until successful decoding of a generation for \gls{fprac}, \gls{sprac}, and \gls{rlnc}.
In comparison to other methods, \gls{fprac} recovers more packets and requires less transmission in various scenarios.
For instance, when $g=100$ and the payload size is \SI{300}{\byte}, \gls{fprac} with $R=10$ recovers $10$ packets compared to $0.9$ recovered packets in \gls{sprac}.
\gls{rlnc} discards partial packets and waits to receive more coded packets.
As a result, the number of total transmissions in the \gls{rlnc} is higher than \gls{fprac} and \gls{sprac}.
For example, when the packet size is \SI{700}{\byte} and $g=100$, the number of transmissions for \gls{fprac} with $R=15$ is $111.6$ compared to $133$ packets in the \gls{rlnc} method.
Furthermore, when $g=100$ and the payload size is \SI{300}{\byte}, \gls{fprac} with $R=15$ requires an average of $105.9$ total transmissions, whereas \gls{sprac} requires $112.3$ packets.

Moreover, the total transmission gap between \gls{fprac} and \gls{sprac} grows from $12.09$ packets for a \SI{300}{\byte} payload to $40.98$ packets for a \SI{700}{\byte} payload.
Nonetheless, transmitting too many redundant packets causes more overhead.
For instance, for \gls{fprac} with a payload size of \SI{300}{\byte}, the total number of required transmissions is $124.3$ and $105.9$ for $R=5$ and $R=10$, respectively.  
	
\section{Conclusion}
\label{sec:conclusion}
Errors are unavoidable in wireless communication, and a portion of packets may become corrupted during transmission over a noisy link.
Even though a significant part of partial packets is error-free, conventional network coding techniques discard them.
We have proposed a packet recovery scheme called \gls{fprac} to exploit partial packets for \gls{rlnc} and \gls{snc}.
The proposed method uses algebraic relation between a group of linearly dependent coded packets to estimate corrupted parts of partial packets.

\gls{fprac} is the first proposed \gls{ppr} that recovers partial packets at the intermediate node and uses the recovered packets for recoding to increase efficiency.
In comparison to existing \gls{acr}-based \gls{ppr} systems, \gls{fprac} reduces the probability of false-positive events and failure during recovery, allowing it to recover more partial packets.

We have compared \gls{fprac} with one of the notable \gls{ppr} schemes for \gls{nc} called \gls{sprac}.
The simulation results reveal that in poor channel conditions, \gls{fprac} improves goodput, and reduces total transmissions as well as completion time for very noisy communication channels.
Moreover, \gls{fprac} reduces completion time, and in an \gls{snc} communication, it improves partial decoding by reducing average decoding delay.

\end{document}